\documentstyle [12pt,epsf]{article}
 
\begin{document}
 
\baselineskip 7.5 mm
 
\begin{flushright}
\begin{tabular}{l}
CERN-TH/98-34 \\
PURD-TH-98-03 \\
hep-ph/9801405
\end{tabular}
\end{flushright}
 
\vspace{12mm}
\begin{center}
 
{\Large \bf
Bound states and resonances in the scalar sector of the MSSM
}
\vspace{18mm}
 
{\large
Alexander Kusenko,}$^{a,}$\footnote{ email address:
Alexander.Kusenko@cern.ch} 
{\large Vadim Kuzmin,}$^{b,}$\footnote{email address: 
kuzmin@ms2.inr.ac.ru} 
{\large and 
Igor I. Tkachev}$^{b,c,}$\footnote{ email address:
tkachev@physics.purdue.edu }  
 
\vspace{6mm}
$^a$Theory Division, CERN, CH-1211 Geneva 23, Switzerland \\
$^b$Institute for Nuclear Research, 60th October Anniversary Prospekt 7a,
Moscow, Russia 117312 \\
$^c$Department of Physics, Purdue University, West Lafayette, IN 47907, USA
\vspace{12mm}

{\bf Abstract}
\end{center}
 
The trilinear couplings of squarks and sleptons to the Higgs bosons can 
give rise to a spectrum of bound states with exotic quantum numbers, for
example, those of a leptoquark.   

\vspace{8mm}
 
\begin{flushleft}
\begin{tabular}{l}
CERN-TH/98-34 \\
PURD-TH-98-03 \\
January, 1998
\end{tabular}
\end{flushleft}

\vfill
 
\pagestyle{empty}
 
\pagebreak
 
\pagestyle{plain}
\pagenumbering{arabic}

The scalar sector of the Minimal Supersymmetric Standard Model (MSSM) 
comprises the Higgs fields and the superpartners of quarks and leptons. 
The gauge interactions of these particles are well understood because the
charges of the scalars, by virtue of supersymmetry, are the same as those of
their fermionic partners.   In addition, the squarks and sleptons can 
have some scalar interactions.  The quartic coupling of the scalar fields are
related to the Yukawa couplings and the gauge couplings by supersymmetry.  
However, the strength of the trilinear interactions is to a large extent
unconstrained, except indirectly, from the requirement of vacuum stability.
It is possible (and, in some models, desirable) that these dimensionful
couplings be large in comparison to the masses of some scalar particles.  
The main focus of our analyses is on such trilinear scalar interactions
because they are effectively ``attractive'' and can cause the appearance of
bound states in the theory.  These bound states can be observed as
resonances in high-energy experiments and can, for example, cause an
increase in the cross-section qualitatively similar to that reported last
year at HERA~\cite{hera}. 

We denote the SU(2)-doublet chiral superfields 
of quarks and leptons as $Q^{\alpha}_{_L}$ and $L^{\alpha}_{_L}$
respectively, and use the same notation for their scalar components. The
corresponding  right-handed SU(2)-singlets are $u_{_R}$, 
$d_{_R}$, and $l_{_R}$.  Here and below the Greek letters stand for the
SU(2) indices, the color SU(3) indices are suppressed.  When relevant, the
additional flavor indices, in Latin characters, will indicate the
generations of (s)quarks and (s)leptons.  

The superpotential of the MSSM includes the following terms 
\begin{equation}
W=\epsilon^{\alpha \beta} [
y_u Q^{\alpha}_{_L} H_2^{\beta} u_{_R} +y_d Q^{\alpha}_{_L} H^{\beta}_1
d_{_R} + y_l L^{\alpha}_{_L} H^{\beta}_2 l_{_R}  
- \mu H_1^{\alpha} H^{\beta}_2] +...,  
\label{sptn}
\end{equation}
and generates, in particular, the trilinear interactions of the form 

\begin{equation}
V_{3,\mu} = 2 y \mu \epsilon^{\alpha \beta} H^{\alpha}_1  Q^{\beta}_{_L}
u_{_R} + {\rm h. c.}   
\end{equation}
that preserve supersymmetry.  In addition, the trilinear terms  

\begin{equation}
V_{3,{_A}} =  A \epsilon^{\alpha \beta} H^{\alpha}_2  Q^{\beta}_{_L}
u_{_R} + {\rm h. c.},  
\label{A}
\end{equation}
which couple the same squark bilinear $Q_{_L} u_{_R}$ to the ``wrong'' 
Higgs, appear in the scalar potential as a consequence 
of supersymmetry breaking.  The potential can be written as follows

\begin{equation}
V=V_{2,{H}}+ V_2 +   V_{3,\mu} +V_{3,{A}} + V_4,  
\end{equation}
where $V_{2,{H}}$ comprises the terms quadratic in the Higgs fields, 
$V_2$ contains the mass terms of the squarks and sleptons,  $V_4$ comprises
the gauge D-terms and the terms of the form $y^2 Q^2 q^2$, 
$y^2 Q^2 H^2$, {\it etc. }

The trilinear coupling can play a crucial role in creating the bound
states and resonances of squarks and sleptons through the exchange of 
the Higgs fields.  For clarity, we will assume
a particular form for the scalar potential that will retain all the
relevant features of the general scalar interaction in the MSSM but will
greatly simplify the
discussion.  First, we make use of a well-known MSSM prediction  that
there is a light neutral Higgs $h^0$.  It is reasonable to neglect the
propagation of heavier Higgs scalars in the ladder diagrams.  
Second, we will assume that the squarks and the sleptons have 
degenerate masses around $m_0$ (this assumption  could be motivated by 
the constraints on the FCNC, although the latter can be satisfied without
the squark degeneracy).  In addition, we assume the equality of the trilinear
couplings for the squarks and sleptons in question.  One can easily generalize
on all these assumptions, none of which is crucial to the main conclusions
of our analysis.  However, the algebraic entanglement involved in tracking a
large number of parameters may unnecessarily complicate the discussion. 
We now want to examine the possibility of a bound state of 
two squarks, two sleptons, or a squark and a slepton, that exchange the
lightest Higgs boson.  The relevant interaction can be described by the
approximate potential written in the flavor-diagonal basis of squarks and
sleptons:  

\begin{equation}
V_a= m_0^2 (|Q^\alpha|^2+|q|^2 +|L^\alpha|^2 +|l|^2) 
+\frac{1}{2} m_{_H}^2 |H^\alpha|^2- A  
\epsilon^{\alpha \beta} H^{\alpha} Q^{\beta} q 
- A \epsilon^{\alpha \beta} H^{\alpha} L^{\beta} l +..., 
\label{aptn}
\end{equation}
The quartic couplings
$V_4$ can be neglected as long as the interaction relevant for the
Bethe-Salpeter equation is dominated by the  large trilinear
terms, the case in which we are mainly interested. 

Theories of the kind described by the potential in equation
(\ref{aptn}) have historically been the testing ground for solving the 
Bethe--Salpeter equation~\cite{wc},  

\begin{equation}
\left [ \left (\frac{1}{2} E + p \right)^2 +m_0^2 \right ]
\left [ \left (\frac{1}{2} E - p \right)^2 + m_0^2 \right ]
\psi (p) = \frac{4 i A^2}{(2\pi)^4}
\int d^4k  \frac{\psi(k)}{(p-k)^2+m_{_H}^2}, 
\label{bs}
\end{equation} 
where $\psi(p)$ is the wave function, $E$ is the bound state energy, and
$m_{_H}$ is the physical mass of the (lightest) Higgs.  The potential in
equation  (\ref{aptn}) can be re-written in the form that matches exactly
the interaction studied in Ref.~\cite{ld1} if one diagonalizes the squark
and slepton bilinears that enter into the cubic terms by a unitary 
transformation.  We can, therefore, use the results of
Refs.~\cite{wc,ld,ld1} for the energy spectrum of the bound states. 

For fixed energy $E$, equation (\ref{bs}) is a Fredholm
equation that has a discrete spectrum of eigenvalues $\lambda \equiv A^2$
that depend on $E$.  In other words, for a given value of the coupling
$\lambda$ one looks for such energy $E$ that makes $\lambda $ an
eigenvalue.  Then the bound state energy $E$ is characterized by a
discrete spectrum. If $m_{_H}=0$ and $\alpha$ is small, the $n$'th bound
state has energy~\cite{wc,ld}   

\begin{equation}
E_n = 2 m_0 \left (1- \frac{\alpha^2}{8 n^2} \right ),  
\label{spectrum}
\end{equation}
where

\begin{equation}
\alpha = \frac{1}{16 \pi} \frac{A^2}{m_0^2}. 
\label{alpha}
\end{equation}

The bound states exist for any value of $\alpha$ if the Higgs field $H$ 
is massless.  In the ladder approximation, the bound state energy approaches
zero at $\alpha = \pi n (n+1) $, which can be used as a semi-quantitative 
reference point for the strength of the attractive interaction. 
The exchange of a scalar field with mass $m_{_H}$ creates a
bound state only if $\alpha > \alpha_{min} \approx 1.68
(m_{_H}/m_0)$~\cite{ld}.

It is interesting to juxtapose these bound states and Q-balls~\cite{qb}. 
Bosons can form a coherent 
state that allows a semiclassical description in terms of
non-topological solitons if the number of particles is sufficiently large.
The connection between sparticle bound states and Q-balls clarifies 
the role of the trilinear terms.  The existence
of small Q-balls~\cite{ak_qb} in the MSSM relies on the requisite
trilinear couplings~\cite{ak_mssm} that make it possible for the energy of
the coherent state with a given baryon or lepton number to be less than the
combined mass of the free particles that carry the same charge.  The same
trilinear term in the scalar potential can be viewed as an attractive
interaction between the constituent squarks that 
exchange the Higgs fields. The latter description is more appropriate in
the few-body limit, where the  semiclassical description of Q-balls breaks
down.   Nevertheless, it is tempting to compare  the expression for the
ground state energy $E_1$ in equation (\ref{spectrum}) to the mass of a
small Q-ball.  One can suspect that a small Q-ball is merely 
an alternative description of some bound states.  In some cases,
such point of view is justified.  For example, a bound state of two leptons
can be thought of as a Q-ball with charge $Q=2$ associated with the lepton 
number U(1)$_{_L}$ symmetry.  A bound state of a slepton and a squark can be 
seen as a U(1)$_{_{B-L}}$ soliton, and so on.  Of course, only a subset of
sparticle bound states can be linked to Q-balls.   The masses of small
Q-balls (those, for which the thin-wall approximation is not valid) for a
potential (\ref{aptn}) with $m_{_H}=m_0=m$ were calculated in 
Ref.~\cite{ak_qb}:  

\begin{equation} 
E_{_Q}= Q m \left [1- \frac{Q^2}{54 S_\psi^2} \left (\frac{A^2}{m^2} 
\right)^2  \right ],
\label{qball}
\end{equation}
where $S_\psi = 4.85$ is a quantity found numerically.  The expression 
(\ref{qball}) was obtained in a semiclassical approximation that becomes
unreliable when $Q \sim 1$.  The energy in equation (\ref{spectrum}) was
calculated~\cite{wc,ld} in the ladder approximation from the Bethe-Salpeter
equation.  Although there is no reason to expect a good agreement between
the two approximations, one of which (\ref{qball}) is pushed beyond its
limit of validity, we notice that for $n=1$ and $Q=2$ they give the same 
dependency on the parameters $A$ and $m$.  In addition, formula 
(\ref{qball}) would give the same quantitative result if constant $S_\psi$ 
assumed a somewhat higher value.

We believe that taking into account the detailed structure
of the mass terms in the MSSM and the inclusion of all degrees of
freedom in the Higgs sector is unlikely to alter one's conclusions with
respect to the existence of the bound states.   Some generalizations of
formula (\ref{spectrum}) to the case of the scalar fields with different
masses and unequal trilinear couplings can be found in~\cite{ld1}.  We
conclude that the squarks and sleptons of the MSSM can form bound states.
The binding energy is determined by the coupling $\alpha$ in equation
(\ref{alpha}).  If the squarks and the sleptons have different masses
$m_1$ and $m_2$, and couple to the light Higgs with the couplings $A_1$ and
$A_2$ respectively, then the relevant coupling is $\tilde{\alpha} =
(1/16\pi) A_1 A_2/ m_1 m_2$~\cite{ld1}. 

Phenomenologically, the trilinear coupling $A$ in equation (\ref{A}) can
be as large as a few TeV.  The upper limit on the value of $A$ comes from
considerations of vacuum stability with respect to tunneling into a
possibly lower color and charge breaking (CCB) minimum in the scalar 
potential~\cite{ccb,kls}.  The strongest limit of this kind exists for the
third generation trilinear coupling, $A_t$, because the tunneling into a
CCB minimum associated with a small Yukawa coupling is
suppressed~\cite{chh}. The empirical formula inferred from the numerical
analyses~\cite{kls} is 

\begin{equation}
(A_t/y_t)^2 +3 \mu^2 < 7.5 (m_{\tilde{t}_{L}}^2+m_{\tilde{t}_{R}}^2),
\label{Alimit}
\end{equation}
where $m_{\tilde{t}_{L}}$ and $m_{\tilde{t}_{R}}$ are the squark masses and
$y_t \approx 1$ is the top Yukawa coupling\footnote{
If one requires that the color and charge conserving vacuum be the global
minimum of the potential, the coefficient 7.5 in equation (\ref{Alimit}) is
replaced by 3~\cite{ccb}.  We note that  $A_t$ in
Refs.~\cite{ccb,kls} differs from ours because we absorbed the Yukawa
coupling into the definition of $A$.}.  Clearly, the ratio
$A_t/m_{\tilde{t}_{R}}$ can be of order 10 if the right-handed stop is very
light.

\begin{figure}
\setlength{\epsfxsize}{3.5in}
\centerline{\epsfbox{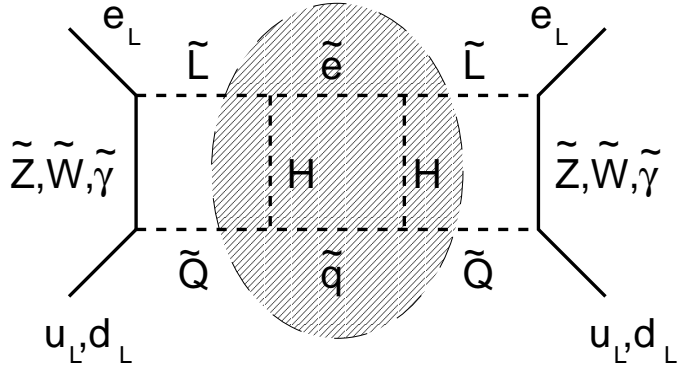}}
\caption{
The scalar ``leptoquark'' bound state can create a resonance at an
electron-proton collider.  The SU(2)-doublets $\tilde{Q}$ and $\tilde{L}$ 
are the first-generation squark and slepton if the gaugino is $\tilde{Z}$
or $\tilde{\gamma}$.  They can be of a different flavor (for example, stop
and $\tilde{\tau}$) for the $\tilde{W}$ exchange.  Only one of many
possible diagrams is shown. 
}
\label{fig_lq}
\end{figure}

While the masses and widths of the bound states may vary, their quantum
numbers are determined by the particle content of the MSSM.  Many of them
can produce 
resonances at present and future experiments.  For example, let us consider
a ``leptoquark'' bound state that can show up as a resonance in an 
electron-proton collider.  The corresponding diagram is shown in
Fig.~\ref{fig_lq}.  
This resonance can be observed through an increase in the cross-section,
much like that reported by HERA experiments~\cite{hera} at high $Q^2$. 
The experimental status of these events remains uncertain but will undoubtedly
be clarified in the near future.  A squark with mass $\sim 200$ GeV and 
with R-parity violating couplings~\cite{squark}  has been proposed as an
explanation of the HERA events.  We emphasize that the leptoquark
resonances can exist in the MSSM with conserved R-parity. They correspond
to the bound states of squarks and sleptons of the type illustrated in
Fig.~\ref{fig_lq}.  Perhaps, the resonances in the 200 GeV mass range can 
account for the HERA events if their coupling to the light fermions is
sufficiently large.  Of course, the parameters of such resonances are
model-dependent.   

\begin{figure}
\setlength{\epsfxsize}{4.5in}
\centerline{\epsfbox{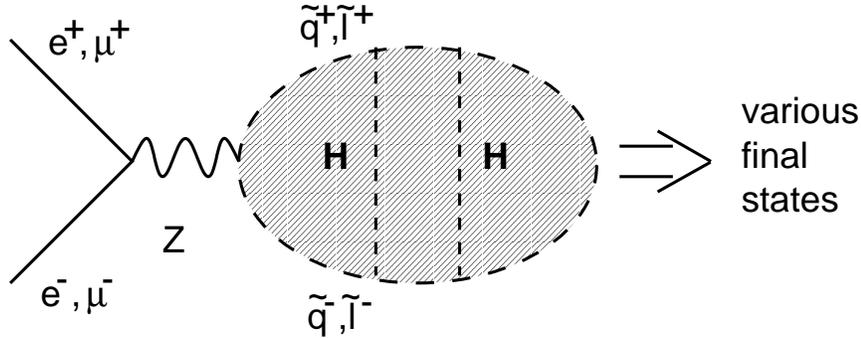}}
\caption{
An example of a resonance at the $e^+e^-$ or $\mu^+\mu^-$ collider. 
}
\label{fig_lepton}
\end{figure}

We expect a variety of resonances to be detectable at a lepton collider,
for instance, at LEP, NLC, or a muon collider (Fig.~\ref{fig_lepton}). 

The two-particle bound states can have two squarks (including a 
``positronium'' state that comprises a squark and an anti-squark), two 
sleptons, or a slepton and a squark.  
Depending on the trilinear terms, the leptoquark resonances 
can have different quantum numbers that correspond to $e^\pm d$, $e^\pm u$,
etc.  In addition, there can be colorless three-particle states bound by
the exchange of the Higgs fields ({\it cf.} Ref.~\cite{ld1}) as well as  
gluons. The states of greatest interest are,
presumably, those that correspond to the large trilinear couplings and
have, therefore, a greater binding energy.  Some of these multi-particle 
states may also show up as resonances.  We leave the details for future
publication.  

In summary, we have shown that large trilinear couplings in the 
scalar sector of the MSSM can give rise to a new family of bound states and
resonances, in particular those with the exotic quantum numbers, that may
be observed in experiment.  

We thank S.~Lola and M.~Shaposhnikov for helpful discussions.  V.~A.~Kuzmin
and I.~I.~Tkachev thank Theory Division at CERN for hospitality.  
I.~I.~Tkachev was supported in part by the U.S. Department of Energy
under Grant DE-FG02-91ER40681 (Task B) and by the National Science
Foundation under Grant PHY-9501458.

\end{document}